\documentclass[twocolumn,english]{revtex4-2}
\usepackage[T1]{fontenc}
\usepackage[latin9]{luainputenc}
\usepackage{babel}
\usepackage{amstext}
\usepackage{graphicx}
\usepackage[pdfusetitle,
 bookmarks=true,bookmarksnumbered=false,bookmarksopen=false,
 breaklinks=false,pdfborder={0 0 1},backref=false,colorlinks=false]
 {hyperref}

\makeatletter

\providecommand{\tabularnewline}{\\}

\makeatother

\begin{document}
\title{2D additive small-world network with distance-dependent interactions}
\author{R. A. Dumer }
\email{rafaeldumer@fisica.ufmt.br}

\affiliation{Programa de Pós-Graduação em Física, Instituto de Física, Universidade
Federal de Mato Grosso, Cuiabá, Brasil.}
\author{M. Godoy}
\email{mgodoy@fisica.ufmt.br}

\affiliation{Programa de Pós-Graduação em Física, Instituto de Física, Universidade
Federal de Mato Grosso, Cuiabá, Brasil.}
\begin{abstract}
In this work, we have employed Monte Carlo calculations to study the
Ising model on a 2D additive small-world network with long-range interactions
depending on the geometric distance between interacting sites. The
network is initially defined by a regular square lattice and with
probability $p$ each site is tested for the possibility of creating
a long-range interaction with any other site that has not yet received
one. Here, we used the specific case where $p=1$, meaning that every
site in the network has one long-range interaction in addition to
the short-range interactions of the regular lattice. These long-range
interactions depend on a power-law form, $J_{ij}=r_{ij}^{-\alpha}$,
with the geometric distance $r_{ij}$ between connected sites $i$
and $j$. In current two-dimensional model, we found that mean-field
critical behavior is observed only at $\alpha=0$. As $\alpha$ increases,
the network size influences the phase transition point of the system,
i.e., indicating a crossover behavior. However, given the two-dimensional
system, we found the critical behavior of the short-range interaction
at $\alpha\approx2$. Thus, the limitation in the number of long-range
interactions compared to the globally coupled model, as well as the
form of the decay of these interactions, prevented us from finding
a regime with finite phase transition points and continuously varying
critical exponents in $0<\alpha<2$.
\end{abstract}
\maketitle

\section{Introduction}

The small-world network model emerges as a proposal to capture the
properties observed in Milgram's social experiment \citep{1}. In
this 1967 experiment, it was found that any person can contact another
with an average of only six intermediaries. Accordingly, a network
that encompasses these properties should have a small average separation
between nodes, compared to the maximum network lenght, as well as
a high local clustering, given the nature of social networks. Two
main network models reproduce these characteristics: the rewired small-world
network (R-SWN) \citep{2} and the additive small-world network (A-SWN)
\citep{3}. In the first case, we start with a regular network, each
node is visited, and with probability $p$, the edges are randomly
rewired, keeping the visited node as one of the endpoints of the edge.
In the case A-SWN, we also start with a regular network, each node
is visited, but with probability $p$, a random edge is created connecting
the visited node with any other node in the network. The difference
between the models lies in the greater simplicity of the A-SWN to
implement and guarantee the small-world phenomenon for a wide range
of $p$, i.e., $p\ge1/N$ with $N$ being the number of nodes in the
network. On the other hand, the R-SWN only maintains the small-world
properties, more specifically the high local clustering, at low values
of $p$.

A generalization of the Ising model \citep{4} can be made by considering
a globally coupled network, which has $N(N-1)/2$ distinct edges.
In this context, the sites of this network are described as magnetic
moments that can take only two values, $\sigma=\pm1$, and the edges
are interpreted as interactions between these magnetic moments, which
can decay with the geometric distance, $r_{ij}$, between connected
sites $i$ and $j$, with $J_{ij}=r_{ij}^{-\alpha}$. In this generalization,
the critical behavior of the system depends on $\alpha$, with three
regimes expected \citep{5,6}: \emph{(i)} mean-field critical exponents
for $0<\alpha<\alpha_{1}$; \emph{(ii)} continuously varying critical
exponents for $\alpha_{1}<\alpha<\alpha_{2}$; and \emph{(iii)} critical
exponents of the model with only short-range interactions for $\alpha>\alpha_{2}$.
Both approximate analytical results and computational simulations
show that $\alpha_{1}=3d/2$ and $\alpha_{2}\approx d+2$, with $d$
being the dimension of the network in question.

In this context, even with a wide range of complex networks already
studied in physical systems \citep{7,8,9,10}, there remains an open
interest in verifying whether the long-range interactions present
in the small-world network also affect the critical behavior of the
Ising model if they depend on the distance between the sites in the
network. The homogeneous interaction model, where $J_{ij}=1$ for
any $r_{ij}$, reveals the critical behavior described by the critical
exponents of the mean-field approximation, both in one \citep{11}
and two dimensions \citep{12,13}. However, by implementing spatial
dependence in $J_{ij}$, the Ising model on the A-SWN in one dimension
\citep{14} only exhibited mean-field critical exponents at $\alpha=0$,
resulting in $\alpha_{1}=\alpha_{2}=0$, because for $\alpha>0$,
the long-range interactions did not significantly contribute to maintaining
the system ordered in the thermodynamic limit. This resulted in a
crossover behavior, where magnetic disorder is reached for any $\alpha>0$
when using sufficiently large networks.

Thus, in the present work, we aim to investigate the critical behavior
of the Ising model on a 2D A-SWN, where the long-range interactions
depend on the geometric distance between connected sites. This is
because the 2D model on a regular lattice exhibits spontaneous magnetization
at $T>0$, which consequently suggests the possibility of finding
the three regimes predicted in the globally coupled model, with $\alpha_{1}\ne\alpha_{2}\ne0$.
To this end, we limit ourselves to the case of $p=1$, where every
site in the network of size $N=L^{2}$ will have a long-range interaction
that depends on the distance between connected sites according to
$J_{ij}=r_{ij}^{-\alpha}$.

This article is organized as follows: In Section \ref{sec:Model},
we present the network, and the dynamic involved in the system. We
also provide details about the Monte Carlo method (MC), the thermodynamic
quantities of interest, and the scaling relations for each of them.
The results are discussed in Section \ref{sec:Results}. And in Section
\ref{sec:Conclusions}, we present the conclusions drawn from the
study.

\section{Model and Method\protect\label{sec:Model}}
\begin{center}
\begin{figure}
\begin{centering}
\includegraphics[scale=0.55]{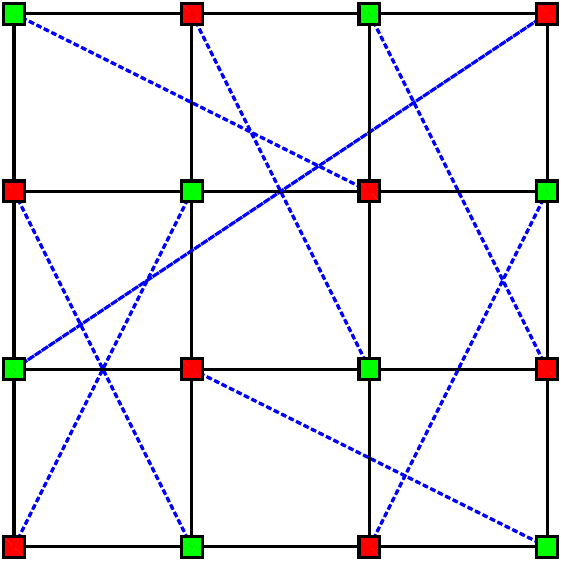}
\par\end{centering}
\caption{{\footnotesize Schematic representation of the network used throughout
the work, with $L=4$ and $N=16$. The black lines indicate short-range
interactions between nearest neighbors sites on the regular lattice,
with interaction strength $J=1$, and the blue dotted line indicates
the long-range interactions added to the network, with interaction
strength $J_{ij}=r_{ij}^{-\alpha}$. The green squares represent one
of the sublattice sites, while the red squares represent the other
sublattice sites, such that the long-range interactions connect the
two sublattice sites. \protect\label{fig:1}}}

\end{figure}
\par\end{center}

\begin{center}
\begin{figure}
\begin{centering}
\includegraphics[scale=0.55]{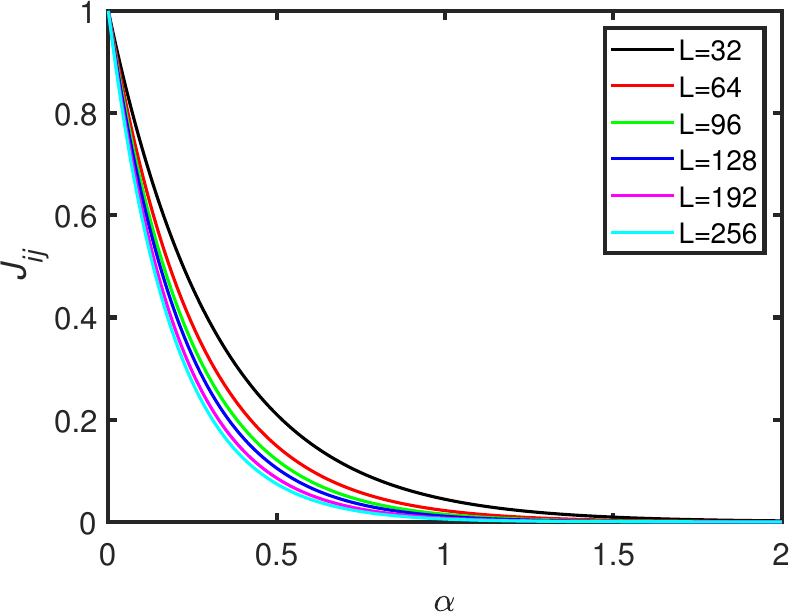}
\par\end{centering}
\caption{{\footnotesize Behavior of $J_{ij}$ as a function of $\alpha$ for
different network sizes, as shown in the figure. The interaction strength
$J_{ij}$ is measured for sites that are separated by a distance equal
to half the maximum distance between two sites on the network, i.e.,
$r_{ij}=\sqrt{2L^{2}}/2$. \protect\label{fig:2}}}
\end{figure}
\par\end{center}

In the ferromagnetic Ising model on the 2D A-SWN, the interaction
energy between spins is defined by the Hamiltonian in the following
form:

\begin{equation}
\mathcal{H}=-\sum_{\left\langle i,j\right\rangle }J\sigma_{i}\sigma_{j}-\sum_{i}J_{ij}\sigma_{i}\sigma_{j},\label{eq:1}
\end{equation}
where $\sigma_{i}=\pm1$, the first sum is taken over the nearest
neighbor spins on the regular square lattice of size $N=L^{2}$, and
the second sum is taken over all the sites in the network. In this
case, we can define $J=1$ and $J_{ij}=r_{ij}^{-\alpha}$, with $r_{ij}$
being the distance between spins $\sigma_{i}$ and $\sigma_{j}$ connected
by the long-range interaction. A schematic representation of a 2D
A-SWN can be seen in Fig. \ref{fig:1}. For simplicity, we decided
to keep $p=1$ throughout the work, meaning all the sites in the network
have a long-range interaction, and these interactions connect two
sublattices, aiming to compare future results with the antiferromagnetic
interaction model. The decay mode of the strength of long-range interactions
is presented in Fig. \ref{fig:2} for different network sizes and
a fixed distance, half the maximum possible length in the network.
Here, we can see the distinction in the interaction strength for small
values of $\alpha$ when we have different network sizes in the system.

We simulate the system specified by the Hamiltonian in Eq. (\ref{eq:1}),
employing MC simulations. We define the initial state with all spins
having the same value, $\sigma_{i}=1$, and always under periodic
boundary conditions. A new spin configuration is generated following
the Markov process: for a given temperature $T$, network size $N$,
and exponent $\alpha$, we randomly select a spin $\sigma_{i}$ e
and alter its state according to the Metropolis prescription \citep{15}:

\begin{equation}
W(\sigma_{i}\to\sigma_{i}^{\prime})=\left\{ \begin{array}{cccc}
e^{\left(-\Delta E_{i}/k_{B}T\right)} & \textrm{if} & \Delta E_{i}>0\\
1 & \textrm{if} & \Delta E_{i}\le0 & ,
\end{array}\right.\label{eq:6}
\end{equation}
where $\Delta E_{i}$ is the change in energy, based in Eq. (\ref{eq:1}),
after flipping the spin $\sigma_{i}$, $k_{B}$ is the Boltzmann constant,
and $T$ the temperature of the system. In summary, a new state is
accepted if $\Delta E_{i}\le0$. However, if $\Delta E>0$ the acceptance
is determined by the probability $\exp\left(-\Delta E_{i}/k_{B}T\right)$,
and it is accepted only if a randomly chosen number $\xi$ uniformly
distributed between zero and one satisfies $\xi\le\exp\left(-\Delta E_{i}/k_{B}T\right)$.
If none of these conditions are satisfied, the state of the system
remains unchanged. 

Repeating the Markov process $N$ times constitutes one Monte Carlo
Step (MCS). We allowed the system to evolve for $10^{5}$ MCS to reach
an equilibrium state, for all network sizes, $(32)^{2}\leq N\leq(256)^{2}$.
To calculate the thermal averages of the quantities of interest, we
conducted an additional $9\times10^{5}$ MCS, and the averages over
the samples were done using $10$ independent samples of the initial
state of the system. The statistical errors were calculated using
the Bootstrap method \citep{16}.

The measured thermodynamic quantities in our simulations are magnetization
per spin $\textrm{m}_{\textrm{L}}$, magnetic susceptibility $\textrm{\ensuremath{\chi}}_{\textrm{L}}$
and reduced fourth-order Binder cumulant $\textrm{U}_{\textrm{L}}$:

\begin{equation}
\textrm{m}_{\textrm{L}}=\frac{1}{N}\left[\left\langle \sum_{i=1}^{N}\sigma_{i}\right\rangle \right],\label{eq:8}
\end{equation}

\begin{equation}
\textrm{\ensuremath{\chi}}_{\textrm{L}}=\frac{N}{k_{B}T}\left[\left\langle \textrm{m}_{\textrm{L}}^{2}\right\rangle -\left\langle \textrm{m}_{\textrm{L}}\right\rangle ^{2}\right],\label{eq:10}
\end{equation}

\begin{equation}
\textrm{U}_{\textrm{L}}=1-\frac{\left[\left\langle \textrm{m}_{\textrm{L}}^{4}\right\rangle \right]}{3\left[\left\langle \textrm{m}_{\textrm{L}}^{2}\right\rangle ^{2}\right]},\label{eq:11}
\end{equation}
where $\left[\ldots\right]$ representing the average over the samples,
and $\left\langle \ldots\right\rangle $ the thermal average over
the MCS in the equilibrium state.

Near the critical temperature $T_{c}$, the equations (\ref{eq:8}),
(\ref{eq:10}) and (\ref{eq:11}) obey the following finite-size scaling
relations \citep{17}:

\begin{equation}
\textrm{m}_{\textrm{L}}=L^{-\beta/\nu}m_{0}(L^{1/\nu}\epsilon),\label{eq:12}
\end{equation}

\begin{equation}
\textrm{\ensuremath{\chi}}_{\textrm{L}}=L^{\gamma/\nu}\chi_{0}(L^{1/\nu}\epsilon),\label{eq:13}
\end{equation}

\begin{equation}
\textrm{U}_{\textrm{L}}^{\prime}=L^{1/\nu}\frac{U_{0}^{\prime}(L^{1/\nu}\epsilon)}{T_{c}},\label{eq:14}
\end{equation}
where $\epsilon=(T-T_{c})/T_{c}$, $\beta$, $\gamma$ and $\nu$
are the critical exponents related the magnetization, susceptibility
and correlation length, respectively. The functions $m_{0}(L^{1/\nu}\epsilon)$,
$\chi_{0}(L^{1/\nu}\epsilon)$ and $U_{0}(L^{1/\nu}\epsilon)$ are
the scaling functions.

\section{Numerical Results and discussion \protect\label{sec:Results}}
\begin{center}
\begin{figure}
\begin{centering}
\includegraphics[scale=0.55]{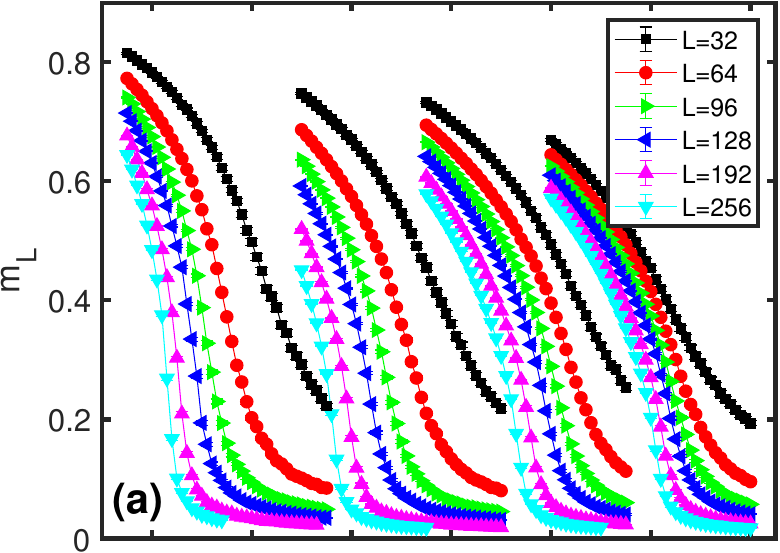}
\par\end{centering}
\begin{centering}
\includegraphics[scale=0.55]{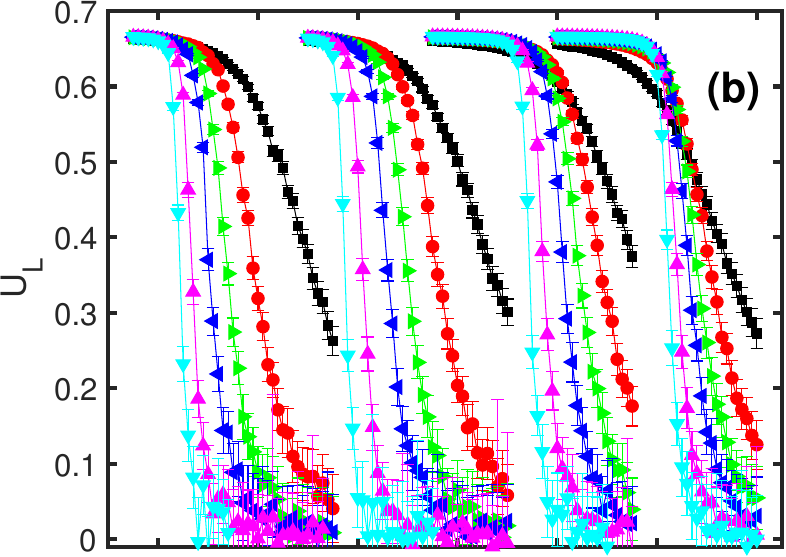}
\par\end{centering}
\begin{centering}
\includegraphics[scale=0.55]{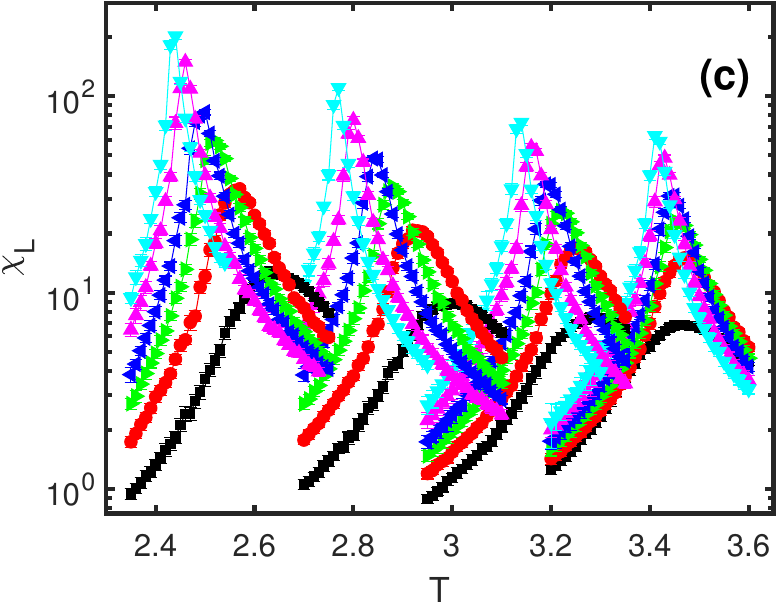}
\par\end{centering}
\caption{{\footnotesize (a) Magnetization $m_{L}$, (b) reduced cumulant $U_{L}$,
and (c) susceptibility $\chi_{L}$ plotted vs temperature $T$ for
various different network sizes $L$, as indicated in panel (a). From
right to left for $\alpha=0.1$, $\alpha=0.2$, $\alpha=0.4$, and
$\alpha=0.8$. \protect\label{fig:3}}}
\end{figure}
\par\end{center}

At low values of $\alpha$, we observed that the system could not
remain ordered in the thermodynamic limit, except for $\alpha=0$
, where we obtained the mean-field critical behavior from the model
with long-range interactions independent of the distance between connected
sites. This crossover behavior, where we can reach critical temperatures
as low as the network size increases, is easily identified with the
curves of thermodynamic quantities for different network sizes. These
curves for different network sizes and low values of $\alpha$ can
be seen in Fig. \ref{fig:3}. In Fig. \ref{fig:3}(a), the curves
of $\textrm{m}_{\textrm{L}}$ can be seen showing the finite-size
behavior, where the larger the network size and the smaller the value
of $\alpha$, the faster the curves decay towards the disordered paramagnetic
phase $P$. The same can be observed in Figs. \ref{fig:3}(b) and
(c) for the curves of $\textrm{U}_{\textrm{L}}$ and $\textrm{\ensuremath{\chi}}_{\textrm{L}}$,
respectively.
\begin{center}
\begin{figure*}
\begin{centering}
\includegraphics[scale=0.55]{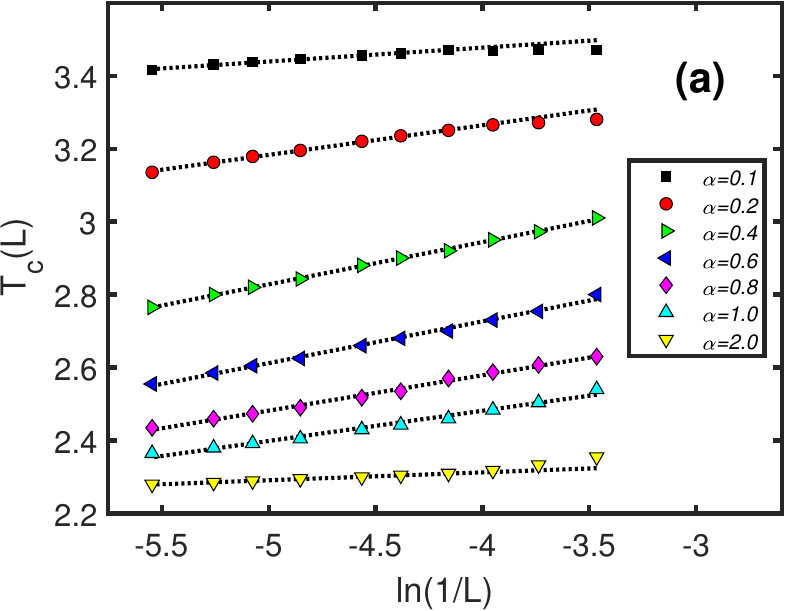}\hspace{0.25cm}\includegraphics[scale=0.55]{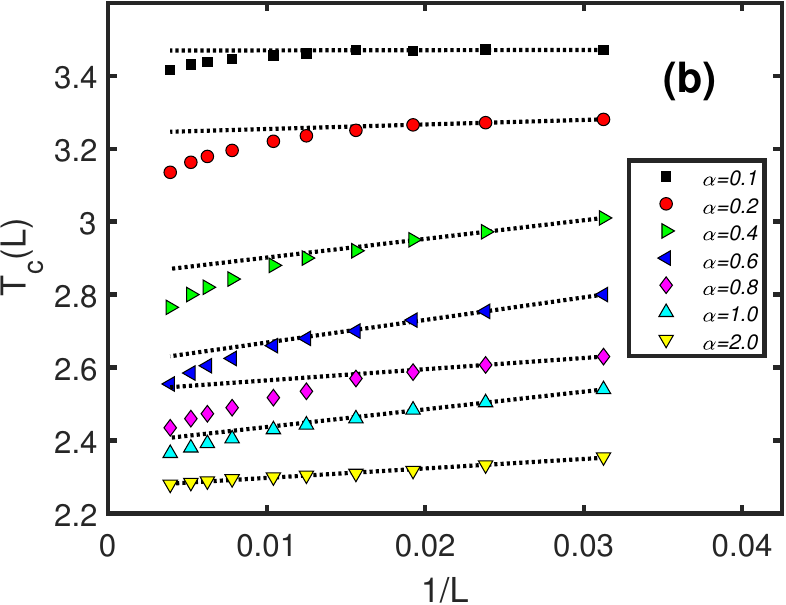}
\par\end{centering}
\caption{{\footnotesize Pseudocritical temperature $T_{c}(L)$ as a function
of (a) $\ln(1/L)$ and (b) $1/L$ for different values $\alpha$,
as indicated in the figures. \protect\label{fig:4}}}
\end{figure*}
\par\end{center}

To confirm the absence of a finite phase transition temperature for
low values of $\alpha$, we can use the finite-size scaling behavior
of $T_{c}(L)$ to extrapolates quite accurately the exact value in
the thermodynamic limit $L\to\infty$. We used the peak of{\footnotesize{}
$\chi_{L}$} for different network sizes $L$, as it indicates the
pseudo-critical temperature $T_{c}(L)$ of the system with linear
size $L$. Therefore, we plotted $T_{c}(L)$ as a function of the
inverse network size $1/L$. Consequently, the linear coefficient
of the linear fit of the points gives us an estimate for the critical
temperature $T_{c}\left(L\to\infty\right)$ in the thermodynamic limit.
This is valid if there is a finite phase transition temperature, as
we have the following scaling relation for finite systems, $T_{c}\left(L\right)=T_{c}\left(L\to\infty\right)\left(1-L^{-1/\nu}\right)$
\citep{18}. In cases where the critical temperature does not have
a linear behavior with the network size, that scaling relation is
not valid, as is our case for low values of $\alpha<2$, as can be
seen in Fig. \ref{fig:4}.
\begin{center}
\begin{table}
\caption{{\footnotesize Values of the coefficients in the linear fit $T_{c}(L)=a+b\ln L$
based on the susceptibility peak, for some values of $\alpha$ in
the crossover regime of the model.\protect\label{tab:1}}}

\centering{}%
\begin{tabular}{ccccccc}
\hline 
$\alpha$ &  &  & $a$ &  &  & $b$\tabularnewline
\hline 
$0.1$ &  &  & $3.6288\pm0.0017$ &  &  & $0.0381\pm0.0073$\tabularnewline
$0.2$ &  &  & $3.5886\pm0.0019$ &  &  & $0.0812\pm0.0094$\tabularnewline
$0.4$ &  &  & $3.4072\pm0.0023$ &  &  & $0.1159\pm0.0176$\tabularnewline
$0.6$ &  &  & $3.1841\pm0.0040$ &  &  & $0.1144\pm0.0243$\tabularnewline
$0.8$ &  &  & $2.9666\pm0.0052$ &  &  & $0.0969\pm0.0238$\tabularnewline
$1.0$ &  &  & $2.8124\pm0.0059$ &  &  & $0.0827\pm0.0251$\tabularnewline
\hline 
\end{tabular}
\end{table}
\par\end{center}

In Fig. \ref{fig:4}(a), we can see that the curves with $\alpha\le1$
better fit the horizontal axis on the logarithmic scale, indicating
a relationship between $T_{c}$ and the network size of the form $T_{c}\left(L\right)=a+b\ln L$.
It is also noticeable that, similar to the case of the one-dimensional
Ising model on an A-SWN network \citep{14}, the smaller the value
of $\alpha$ the larger the network size needed to observe the correct
behavior of $T_{c}$ as a function of $L$. Here, the logarithmic
behavior was verified even for $\alpha=0.02$ with network size $L=768$,
and we extended this result to smaller values of $\alpha$, having
a finite phase transition temperature and mean-field critical behavior
only at $\alpha=0$ \citep{12,13}. The values of the coefficients
of the curves in Fig. \ref{fig:4}(a) that exhibit logarithmic behavior
can be found in Tab. \ref{tab:1}.
\begin{center}
\begin{figure}
\begin{centering}
\includegraphics[scale=0.55]{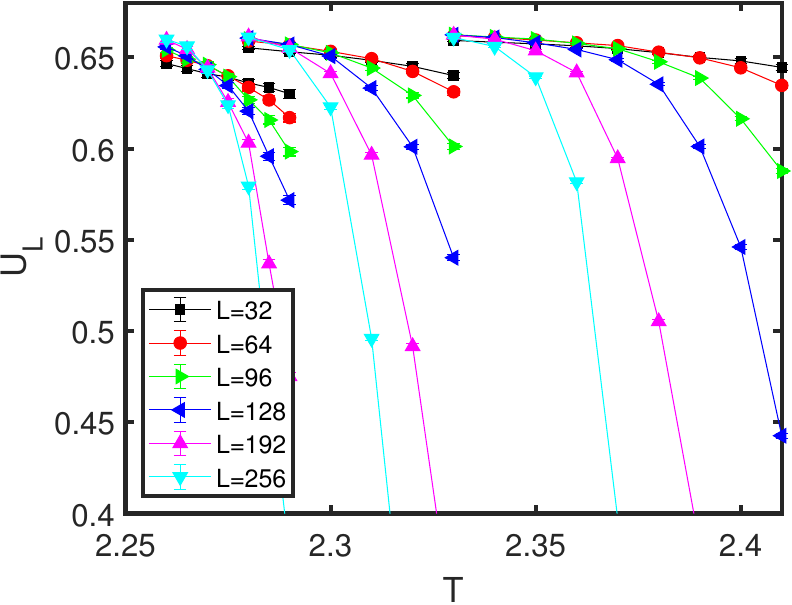}
\par\end{centering}
\caption{{\footnotesize Binder cumulante $U_{L}$ plotted vs temperature $T$
for various different network sizes $L$, as indicated in the figure.
From right to left for $\alpha=1.0$, $\alpha=1.3$, and $\alpha=1.7$.
\protect\label{fig:5}}}
\end{figure}
\par\end{center}

The same curves from Fig. \ref{fig:4}(a) are exhibited on a plot
with a linear horizontal axis (see Fig. \ref{fig:4}(b)), highlighting
the incompatibility of the scaling, which is only valid for small
network sizes at low $\alpha$ values. However, for $\alpha=2$ ,
we have verified that all network sizes are compatible with the linear
scaling relationship of $T_{c}$ as a function of $L$, indicating
a finite critical phase transition temperature.
\begin{center}
\begin{figure}
\begin{centering}
\includegraphics[scale=0.55]{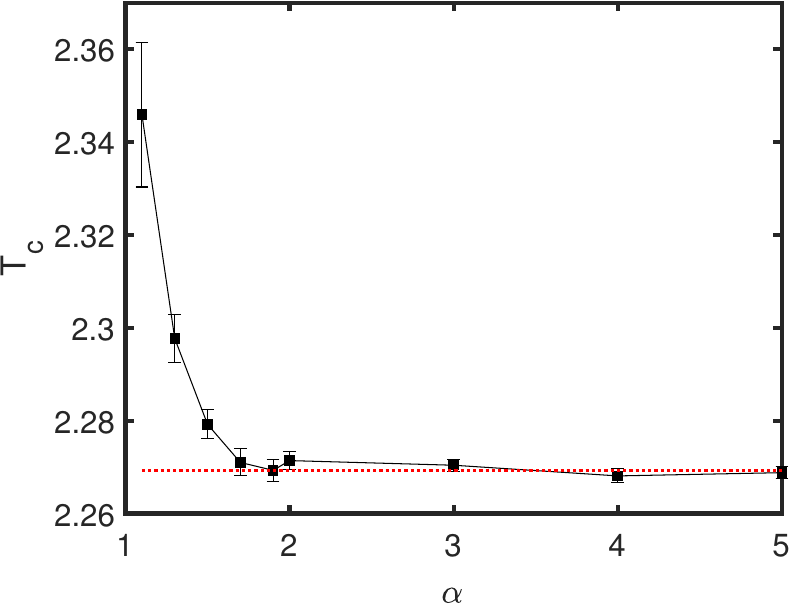}
\par\end{centering}
\caption{{\footnotesize Critical temperature $T_{c}$ as a function of $\alpha$
near the transition point for the regime where the critical behavior
of the system is described by the Ising model with short-range interactions.
The red dotted line is a reference for the exact value $T_{c}$ of
the 2D Ising model on a regular lattice. \protect\label{fig:6}}}
\end{figure}
\par\end{center}

Returning to an analysis of Fig. \ref{fig:3}(b), we can observe a
qualitative behavior indicating a transition between regimes. For
$\alpha=0.1$, the Binder cumulant curves for different network sizes
are closer to each other, even though there is no single crossing
point. As $\alpha$ increases, these curves for different $L$ begin
to diverge more significantly. Thus, for $\alpha=0$, there is a finite
critical temperature, but for $\alpha\ne0$, the system enters a crossover
regime, which becomes increasingly pronounced as $\alpha$ increases,
as shown by the slopes in Tab. \ref{tab:1}. However, observing the
slopes in Tab. \ref{tab:1}, we can see that, they start to decrease
above $\alpha=0.4$ suggesting the onset of a new finite temperature
phase transition regime. This decreasing trend in the slopes continues
up to around $\alpha=2$, as shown in Fig. \ref{fig:5}. In this figure,
we see that as $\alpha$ approaches $2$, the Binder cumulant curves
get closer together, although they still do not exhibit a single crossing
point.
\begin{center}
\begin{figure}
\begin{centering}
\includegraphics[scale=0.55]{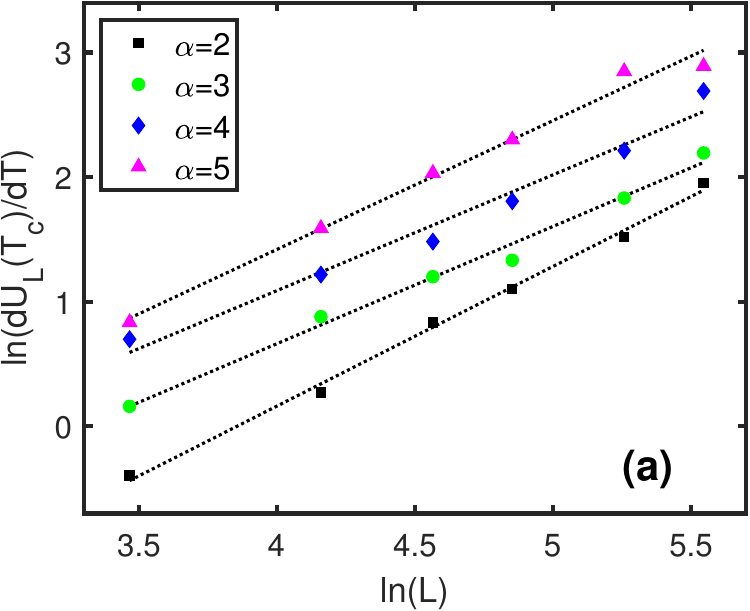}
\par\end{centering}
\begin{centering}
\includegraphics[scale=0.55]{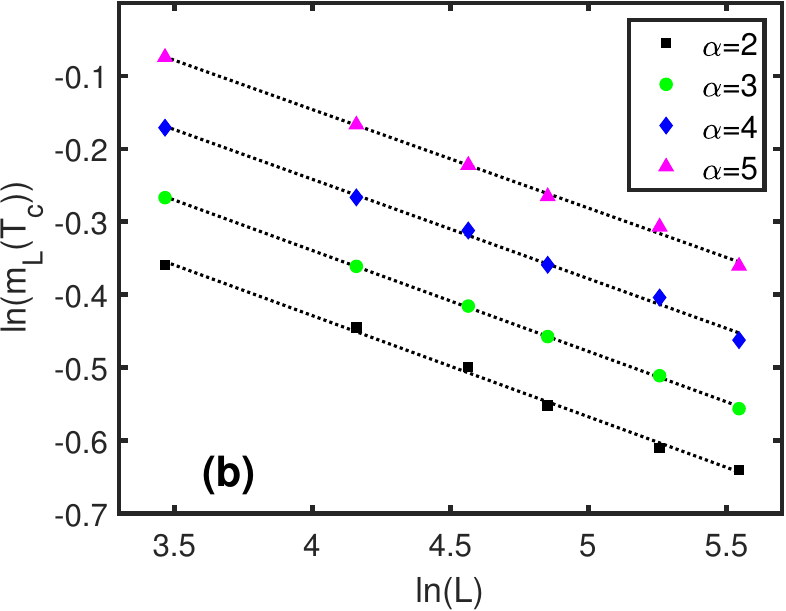}
\par\end{centering}
\begin{centering}
\includegraphics[scale=0.55]{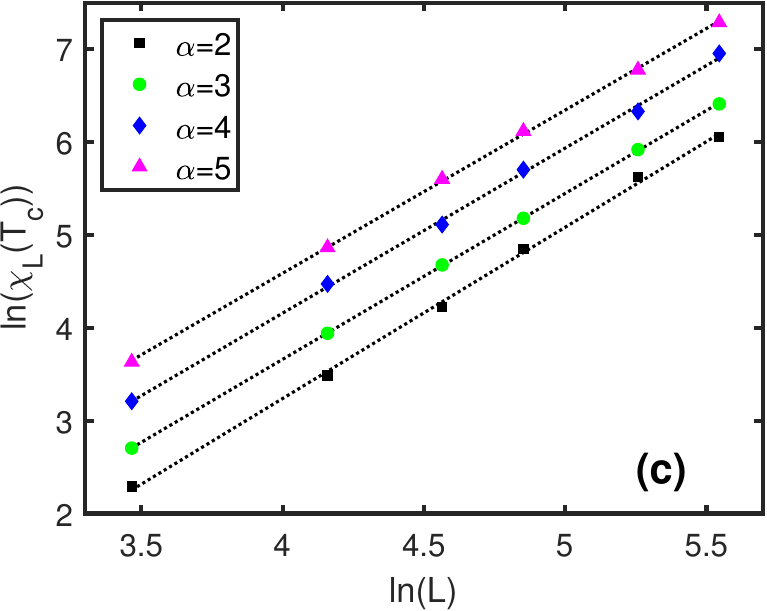}
\par\end{centering}
\caption{{\footnotesize Log-log plot of (a) $m_{L}$, (b) $U_{L}$, and (c)
$\chi_{L}$ near the critical point as a function of $L$, yielding
the critical exponent ratios. In (a) the slope gives $1/\nu$, (b)
$-\beta/\nu$, and (c) $\gamma/\nu$. The curves for different values
of $\alpha$ as presented in the figures. The linear coefficients
of the curves were adjusted for better visualization, and the error
bars are smaller than the size of the symbols. The exponent ratios
can be found in Tab. \ref{tab:2}. \protect\label{fig:7}}}
\end{figure}
\par\end{center}

\begin{center}
\begin{table}
\caption{{\footnotesize Table of critical exponent ratios obtained from the
linear fit of thermodynamic quantities at the critical point for some
values of $\alpha$ in the short-range interaction regime of the model.
The $T_{c}(L\rightarrow\infty)$ was estimated by extrapolation of
the susceptibility peaks.\protect\label{tab:2}}}

\centering{}%
\begin{tabular}{ccccccccc}
\hline 
$\alpha$ &  & $-\beta/\nu$ &  & $\gamma/\nu$ &  & $1/\nu$ &  & $T_{c}(L\rightarrow\infty)$\tabularnewline
\hline 
$2$ &  & $0.139\pm0.01$ &  & $1.84\pm0.11$ &  & $1.12\pm0.12$ &  & $2.271\pm0.0020$\tabularnewline
$3$ &  & $0.138\pm0.01$ &  & $1.79\pm0.04$ &  & $0.94\pm0.10$ &  & $2.270\pm0.0013$\tabularnewline
$4$ &  & $0.136\pm0.02$ &  & $1.78\pm0.13$ &  & $0.93\pm0.13$ &  & $2.268\pm0.0015$\tabularnewline
$5$ &  & $0.135\pm0.01$ &  & $1.76\pm0.06$ &  & $1.03\pm0.11$ &  & $2.269\pm0.0013$\tabularnewline
\hline 
\end{tabular}
\end{table}
\par\end{center}

The entry point into the new finite temperature phase transition regime
was further identified by fitting $T_{c}$ as a linear function of
$L^{-1}$. For this, we selected the value of $\alpha$ where we had
an estimate of $T_{c}$ with an error less than $1\%$. This occurs
near $\alpha=2$, and the behavior of $T_{c}$ as a function of $\alpha$
can be seen in Fig. \ref{fig:6}. In this case, we observe that for
values above $\alpha=2$, there are no significant changes in the
critical temperature, and it is consistent with that obtained for
the Ising model in two dimensions on a regular lattice, $T_{c}=2/\ln\left(1+\sqrt{2}\right)$\citep{19}.
The estimates of $T_{c}$ for $\alpha\ge2$ presented in Fig. \ref{fig:6}
can be found in Tab. \ref{tab:2}.
\begin{center}
\begin{figure*}
\begin{centering}
\includegraphics[scale=0.55]{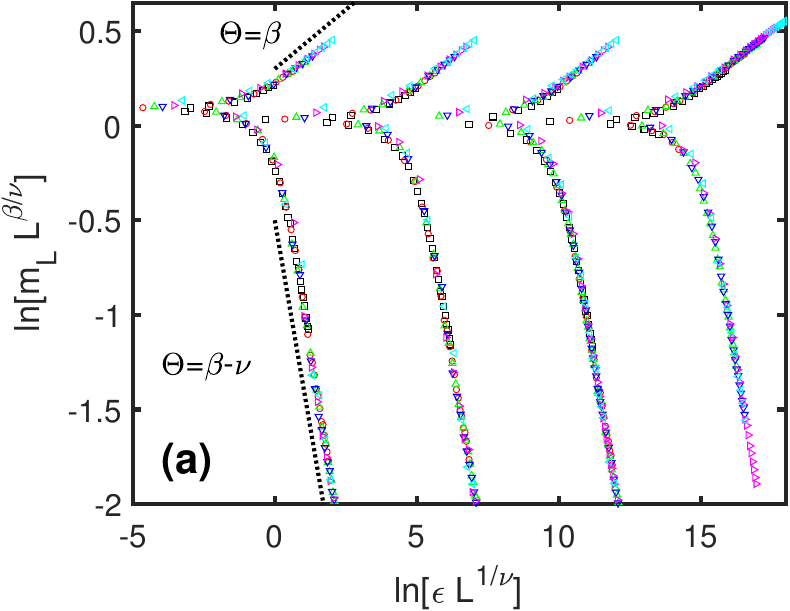}\hspace{0.25cm}\includegraphics[scale=0.55]{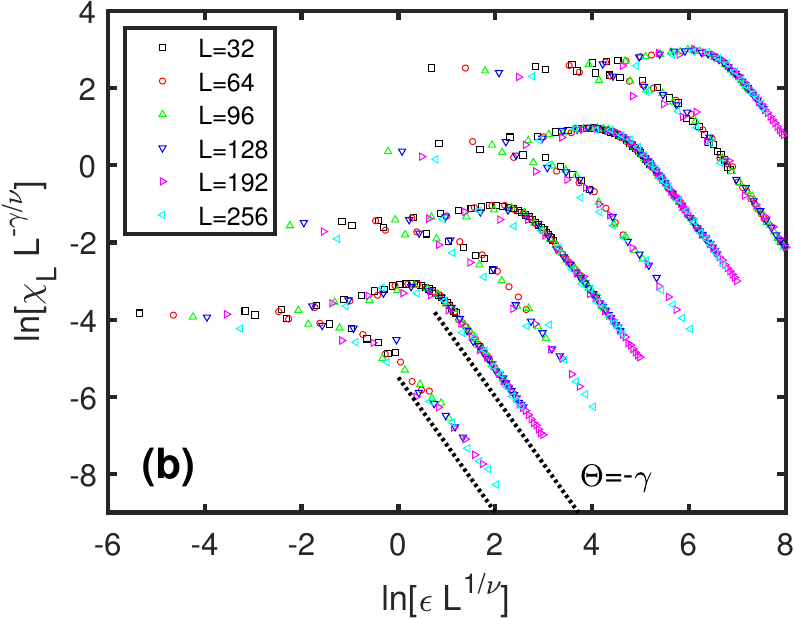}
\par\end{centering}
\caption{{\footnotesize Data collapse into universal scaling functions (a) $m_{L}$
and (b) $\chi_{L}$ for different network sizes $L$, as displayed
in panel (b). (a) From left to right, the collapse is plotted for
$\alpha=2$, $\alpha=3$, $\alpha=4$, and $\alpha=5$, with the curves
shifted horizontally by $5(\alpha-2)$ for better visualization. (b)
From bottom to top, the collapse is plotted for $\alpha=2$, $\alpha=3$,
$\alpha=4$, e $\alpha=5$, with the curves shifted by $2(\alpha-2)$
both vertically and horizontally for better visualization. For all
these collapses were used $\beta=0.125$, $\gamma=1.75$, $\nu=1.0$,
and the exact value $T_{c}$ of the Ising model on a regular 2D lattice.
The dotted lines indicate the asymptotic behavior of the curves with
slope $\Theta$. \protect\label{fig:8}}}
\end{figure*}
\par\end{center}

In order to characterize the finite temperature regime for $\alpha\ge2$
and to determine if, as predicted, it corresponds to the regime of
the Ising model with short-range interactions, we must calculate the
critical exponents of the system. The critical exponents were obtained
using the scaling relations in Eqs \ref{eq:12}, \ref{eq:13}, and
\ref{eq:14}. By employing these relations, we can derive the exponents
from the values of the relevant thermodynamic quantities near the
critical point. This is achieved by log-log plot of these quantities
as a function of the linear size of the network. The slope of the
linear fit to the data points then yields the critical exponent ratios.

Fig. \ref{fig:7}(a) displays the value of the derivative of the Binder
cumulant near the critical point as a function of the linear size
of the network on a log-log plot. Based on the scaling relation in
Eq. \ref{eq:14}, the slope of the linear fit to the points provides
an estimate for the ratio $1/\nu$, which includes the exponent related
to the system's correlation length. Similarly, in Fig. \ref{fig:7}(b),
with the value of the magnetization near the critical point, the linear
fit to the points, based on the scaling relation in Eq. \ref{eq:12},
yields an estimate for the ratio $-\beta/\nu$, which includes the
exponent related to the system's order parameter. In Fig. \ref{fig:7}(c),
the magnetic susceptibility at the critical point, using the scaling
relation in Eq. \ref{eq:13}, allows us to estimate the ratio $\gamma/\nu$.
These estimates can be found in Tab. \ref{tab:2}, and, with the appropriate
errors, they align with the critical exponents of the Ising model
on a regular square lattice, which are $\beta=1/8$, $\gamma=7/4$
and $\nu=1$, confirmed both analytically and through Monte Carlo
simulations \citep{20}. 

The universal behavior for each scaling function, where different
system sizes all collapse in one line, indicates that the critical
exponents obtained for the system present reasonable precision. Therefore,
another method to find the critical exponents of the system, though
it will be used primarily here to verify the validity of the Ising
model exponents in our system, is through data collapse. This method
is also based on the scaling relations Eqs. \ref{eq:12}, \ref{eq:13},
and \ref{eq:14}, and allows us to identify the form of the scaling
function contained in these relations. If the correct critical exponents
are applied to these scaling relations, for curves of different network
sizes, we obtain a single curve that represents the scaling function
of the system. Thus, verify if the critical exponents of the Ising
model on a regular square lattice are applicable to our system, we
have plotted $\textrm{m}_{\textrm{L}}L^{\beta/\nu}$ and $\textrm{\ensuremath{\chi}}_{\textrm{L}}L^{-\gamma/\nu}$
as a function of $L^{1/\nu}\epsilon$, with the exponents $\beta=1/8$,
$\gamma=7/4$, and $\nu=1$. As a result, we obtained a single curve,
which is the scaling function of the respective thermodynamic quantity.

The data collapse for $\textrm{m}_{\textrm{L}}$ and $\textrm{\ensuremath{\chi}}_{\textrm{L}}$
can be found in Fig. \ref{fig:8}, showing that the system exhibits
the behavior of the Ising model with short-range interactions, as
a single curve is obtained for each value of $\alpha$ across different
values of $L$. These plots use logarithmic scaling on the axes to
provide additional confirmation of the exponents of the system through
the asymptotic behavior of the curves away from $\epsilon=0$. The
slope $\Theta=\beta$ is expected for $\epsilon<0$ and $\Theta=\beta-\nu$
for $\epsilon>0$ in the higher-order terms of $m_{0}$ from Eq. \ref{eq:12},
and $\Theta=-\gamma$ in the higher-order terms of $\chi_{0}$ from
Eq. \ref{eq:13}.

\section{Conclusions\protect\label{sec:Conclusions}}

In this work, we studied the Ising model on a 2D A-SWN, where the
long-range interactions $J_{ij}$ depend on the geometric distance
$r_{ij}$ between connected sites $i$ and $j$, such as $J_{ij}=r_{ij}^{-\alpha}$.
We found that the mean-field critical behavior is achieved only at
$\alpha=0$. For $\alpha>0$ , the system enters a crossover regime
where no magnetic ordering is observed in the thermodynamic limit
$L\rightarrow\infty$, as critical temperatures decrease with increasing
network size. However, for $\alpha\ge2$, the system maintains magnetic
ordering due to the diminished influence of long-range interactions,
reaching the regime where the critical behavior is governed by the
short-range interaction Ising model. Compared to the globally coupled
model, which has $N(N-1)/2$ edges, does not exhibit the crossover
behavior observed in the A-SWN, in both one and two dimensions, and
achieves mean-field critical behavior at $\alpha\ne0$. This difference
can be attributed to the limited number of long-range interactions
in the A-SWN, with only $N$ edges. Analyzing $J_{ij}$ as a function
of $\alpha$ in Fig. \ref{fig:2}, we observed that for $0<\alpha<2$,
the different network sizes show different decay rates of long-range
interaction strength. This, combined with the limited number of interactions,
may also explain the emergence of the crossover regime in the system.
However, further studies are required to verify this assertion, where
$J_{ij}$ as a function of $\alpha$ should exhibit the same behavior
for any network size.

\begin{acknowledgments}
This work has been supported by the Conselho Nacional de Desenvolvimento
Científico e Tecnológico (CNPq), Brazil (Process No. 140141/2024-3).
\end{acknowledgments}

\end{document}